\documentclass[a4paper]{article}

\usepackage[utf8]{inputenc}
\usepackage{microtype}
\usepackage{amsmath}
\usepackage{amssymb}
\usepackage{natbib}
\usepackage{arydshln}
\usepackage{graphicx}
\usepackage{authblk}
\hypersetup{colorlinks=true,allcolors=black}

\title{Detecting SNPs with interactive effects \\ on a quantitative trait}

\author{Armin Rauschenberger$^{1}$, Ren\'{e}e X Menezes$^{1}$, Mark A van de Wiel$^{1,2}$, Natasja M van Schoor$^{1}$, and Marianne A Jonker$^{1,3,}$\thanks{marianne.jonker@radboudumc.nl}}

\affil{\small $^{1}$Department of Epidemiology and Biostatistics, VU University Medical Center, Amsterdam, $^{2}$Department of Mathematics, VU University, Amsterdam, $^{3}$Department for Health Evidence, Radboud University Medical Center, Nijmegen}

\date{}

\graphicspath{{./figures/}}

\begin{document}

\maketitle

\begin{abstract}
Here we propose a test to detect effects of single nucleotide polymorphisms~(\textsc{snp}s) on a quantitative trait. Significant \textsc{snp}-\textsc{snp} interactions are more difficult to detect than significant \textsc{snp}s, partly due to the massive amount of \textsc{snp}-\textsc{snp} combinations. We propose to move away from testing interaction terms, and move towards testing whether an individual \textsc{snp} is involved in any interaction. This reduces the multiple testing burden to one test per \textsc{snp}, and allows for interactions with unobserved factors. Analysing one \textsc{snp} at a time, we split the individuals into two groups, based on the number of minor alleles. If the quantitative trait differs in mean between the two groups, the \textsc{snp} has a main effect. If the quantitative trait differs in distribution between \textit{some} individuals in one group and \textit{all} other individuals, it possibly has an interactive effect. We propose a mixture test to detect both types of effects. Implicitly, the membership probabilities may suggest potential interacting variables. Analysing simulated and experimental data, we show that the proposed test is statistically powerful, maintains the type~I error rate, and detects meaningful signals. The R package \href{https://doi.org/10.18129/B9.bioc.semisup}{\texttt{semisup}} is available from Bioconductor.
\end{abstract}

\section{Background}

Many diseases are caused by the genotype or by genotype-environment interactions, ranging from single-gene to complex genetic disorders. Although interactions may cause the missing heritability problem \citep{Zuk2012}, researchers often detect \textsc{snp}s strongly associated with a quantitative trait, but seldom detect significant \textsc{snp}-\textsc{snp} or \textsc{snp}-environment interactions. This is partly due to large-scale multiple testing and the ensuing multiple testing correction. In this methodological study, we propose a powerful genome-wide testing procedure for detecting \textsc{snp}s with interactive effects, which avoids this problem.

We are interested in testing for association between a quantitative trait and numerous \textsc{snp}s. Given that the setting is high-dimensional, a linear model including all \textsc{snp}s, let alone their interactions, would not have all parameters identifiable. A common approach is to analyse \textsc{snp}s one-by-one, ignoring the other \textsc{snp}s, but potentially accounting for some control variables. Then each \textsc{snp} requires a separate regression model.

In principle, we could proceed similarly for analysing interactions of \textsc{snp}s, but the massive number of combinations makes this impractical. As few as $100$~\textsc{snp}s lead to approximately ${5\,000}$~pairs, ${200\,000}$~triplets and ${4\,000\,000}$~quadruplets of \textsc{snp}s. Above all, the human genome contains several million \textsc{snp}s. Usually, the large number of tests will render computation prohibitively expensive, and multiple testing correction will wipe out all significance.

These problems could be alleviated by focussing on \textsc{snp}s with significant main effects \citep{Kooperberg2008}. However, \textsc{snp}s with weak or without main effects can still have an interaction effect \citep{Culverhouse2002}. Such a \textsc{snp} might suppress or activate the effect of an environmental factor. Limiting the order of interactions would decrease the number of tests, but the complexity of biological pathways might break through such a threshold \citep{Taylor2015}. Furthermore, pairwise testing tends to detect \textsc{snp}s with few strong interactions, having less power to detect \textsc{snp}s with many weak interactions. Most epistasis detection methods are only applicable to binary traits, but a recurrent idea is to restrict the search space, through feature selection or feature extraction \citep{VanSteen2011}.

The centrepiece of the proposed testing procedure is the semi-supervised mixture model \citep{Zhu2009}. Analysing one \textsc{snp} at a time, we test whether it has any effect, be it a main effect, an interaction with another \textsc{snp}, or an interaction with another factor. Ignoring all potential interacting variables, we do not construct any interaction terms. Compared to testing all combinations of \textsc{snp}s separately, this approach drastically decreases the number of tests. After detecting \textsc{snp}s that are susceptible to interaction, we could use regression analysis to test specific interaction terms.

In this paper, we implement the mixture test for quantitative traits that follow Gaussian or negative binomial distributions, but extensions to many other distributions should be straightforward. After presenting the semi-supervised mixture test, we first show by simulation that it is statistically powerful and maintains the type~I error rate, and then apply it to detect quantitative trait loci. Results from simulations and applications suggest that the semi-supervised mixture test and classical tests complement each other. The former is more powerful at detecting partial shifts, which are possibly caused by interactions, and the latter are more powerful at detecting complete shifts.

\section{Methods}

\subsection{Main and interactive effects}

In regression analysis, we could combine two or more \textsc{snp}s to an interaction term, and then test its effect on the quantitative trait. To avoid a combinatorial explosion, we want to test whether a \textsc{snp} is involved in any interaction, without constructing interaction terms.

Each individual has either zero, one, or two minor alleles at any \textsc{snp}. We define zero minor alleles as the standard, and one or two minor alleles as the modification, but other options are possible (Section~\ref{discussion}). Analysing one \textsc{snp} at a time, we split the individuals into two groups, one without and one with the modification. The modification does not affect the quantitative trait of individuals without the modification, but may affect the quantitative trait of individuals with the modification.

\begin{figure}[!htbp]
\centerline{\includegraphics[width=1.0\textwidth]{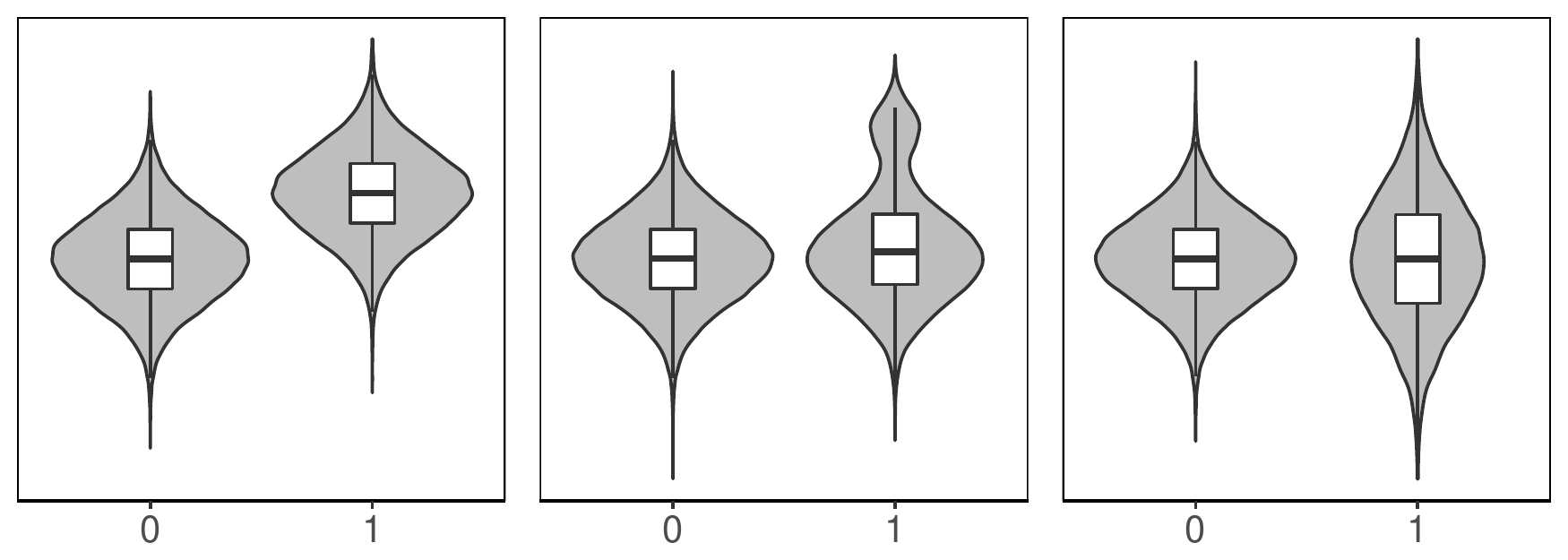}}
\caption{Violin plots of a quantitative trait for individuals without~($0$) and with~($1$) the modification. A \textsc{snp} with a main effect may lead to a complete~(left) or partial~(centre) mean shift. And a \textsc{snp} with an interactive effect may lead to a partial mean shift~(centre) or a variance shift~(right).}
\label{fig:violin}
\end{figure}

If the \textsc{snp} has no effect, the quantitative trait only differs by chance between the two groups. If the \textsc{snp} has a main effect, we might observe a complete or partial mean shift. And if the \textsc{snp} is involved in an interaction, we might observe a partial mean shift or a variance shift. Figure~\ref{fig:violin} shows these three types of shifts. If we compared the means between the two groups, we would detect some main effects, but miss many other effects.

This problem is semi-supervised, as we know one group is unaffected, but we do not know which individuals in the other group are affected. Conveniently, the semi-supervised mixture model is directed towards main and interactive effects. It allows us to test whether the modification affects at least one individual. Consequently, we merely fit and test one model per \textsc{snp}.

\subsection{Semi-supervised mixture model}

We analyse one \textsc{snp} at a time. The numerical variable $\boldsymbol{Y}={(Y_1,\ldots,Y_n)^T}$ represents the quantitative trait, the binary variable $\boldsymbol{X}={(X_1,\ldots,X_n)^T}$ indicates which individuals have the modification, and the binary variable $\boldsymbol{Z}={(Z_1,\ldots,Z_n)^T}$ indicates whether the modification affects the quantitative trait. We index the individuals by ${i=1,\ldots,n}$. Individuals without the modification (${X_i=0}$) are unaffected (${Z_i=\mathrm{A}}$), and those with the modification (${X_i=1}$) are either unaffected (${Z_i=\mathrm{A}}$) or affected (${Z_i=\mathrm{B}}$). We call $\boldsymbol{Y}$ the \textit{observations}, and $\boldsymbol{Z}$ the \textit{class labels}.

Any individual~$i$ belongs either to class A or~B, indicated by ${Z_i=\mathrm{A}}$ and ${Z_i=\mathrm{B}}$. We assume the observations $Y_i$ are independent and come from the discrete or continuous probability distribution~$F$:
\begin{equation}
\begin{split}
Y_i|(Z_i=\mathrm{A}) \sim F(\boldsymbol{\cdot},\boldsymbol{\theta_a}), \\
Y_i|(Z_i=\mathrm{B}) \sim F(\boldsymbol{\cdot},\boldsymbol{\theta_b}),
\end{split}
\end{equation}
where $\boldsymbol{\theta_a}$ and $\boldsymbol{\theta_b}$ are the population parameters. We denote the probability mass or density function for class~A by ${f(\boldsymbol{\cdot},\boldsymbol{\theta_a})}$, and the one for class~B by ${f(\boldsymbol{\cdot},\boldsymbol{\theta_b})}$.
 
Some class labels are observed, and some class labels are missing. If $X_i$ equals zero, $Z_i$ is observed ($Y_i$ is labelled). And if $X_i$ equals one, $Z_i$ is missing ($Y_i$ is unlabelled). By assumption, all labelled observations are in class~A, and unlabelled observations are in class A or~B. Table~\ref{tab:variables} gives an overview of the variables.

\begin{table}[!htbp]
\centering
\begin{tabular}{l|ccccc:ccccc}
& $1$ & $2$ & $3$ & \ldots & $s$ & $s+1$ & $s+2$ & $s+3$ & \ldots & $n=s+u$ \\ \hline
$\boldsymbol{Y}$ & $y_1$ & $y_2$ & $y_3$ & \ldots & $y_{s}$ & $y_{s+1}$ & $y_{s+2}$ & $y_{s+3}$ & \ldots & $y_{s+u}$ \\
$\boldsymbol{X}$ & $0$ & $0$ & $0$ & \ldots & $0$ & $1$ & $1$ & $1$ & \ldots & $1$ \\
$\boldsymbol{Z}$ & A & A & A & \ldots & A & A/B & A/B & A/B & \ldots & A/B
\end{tabular}
\caption{$\boldsymbol{Y}$~is the numerical variable, $\boldsymbol{X}$~indicates the group, and $\boldsymbol{Z}$~indicates the class. The first $s$ observations are labelled, but the last $u$ observations are unlabelled.}
\label{tab:variables}
\end{table}

\subsection{Model fitting}

A random unlabelled observation (${X_i=1}$) belongs to class~A with the unknown probability~$1-\tau$, and to class~B with the unknown probability~$\tau$. Mathematically speaking we have $1-\tau={\mathbb{P}(Z_i=\mathrm{A}|X_i=1)}$ and $\tau={\mathbb{P}(Z_i=\mathrm{B}|X_i=1)}$. We want to estimate the population parameters $\boldsymbol{\theta_a}$ and~$\boldsymbol{\theta_b}$, and the mixing proportion $\tau$ by the maximum likelihood method. The log-likelihood function equals
\begin{equation}
\begin{split}
\log L(\boldsymbol{\theta_a},\boldsymbol{\theta_b},\tau|\boldsymbol{y},\boldsymbol{x}) =&~ \sum\limits_{i=1}^n \Bigg[
(1-x_i) \log f(y_i,\boldsymbol{\theta_a}) \\
&~+~ x_i \log \bigg\{ (1-\tau) f(y_i,\boldsymbol{\theta_a}) + \tau f(y_i,\boldsymbol{\theta_b}) \bigg\}
\Bigg].
\label{eq:incomplete}
\end{split}
\end{equation}
This maximisation problem has no explicit solution. Because the semi-supervised mixture model depends on the missing class labels, we use the expectation-maximisation~(\textsc{em}) algorithm to maximise the likelihood \citep{Dempster1977} (Appendix Section~B). Next to the parameter estimates, the \textsc{em} algorithm also returns the membership probabilities $\boldsymbol{a}={(a_1,\ldots,a_n)^T}$ and $\boldsymbol{b}={(b_1,\ldots,b_n)^T}$. Given the observations, the observed class labels, and the parameter estimates, individual~$i$ belongs to class~A with probability~$a_i$, and to class~B with probability~$b_i$. Excluding the labelled observations, we can correlate the membership probabilities to other genomic or environmental variables, and identify potential interacting variables.

\subsection{Mixture test}

We want to test whether a two-component mixture model fits significantly better to the data than a single-component mixture model. This is equivalent to testing the null hypothesis ${H_0: \tau = 0}$ against the alternative hypothesis ${H_1: \tau > 0}$.

Under the null hypothesis, the unlabelled observations are drawn from ${F(\boldsymbol{\cdot},\boldsymbol{\theta_a})}$, and under the alternative hypothesis they are drawn from ${F(\boldsymbol{\cdot},\boldsymbol{\theta_a})}$ and ${F(\boldsymbol{\cdot},\boldsymbol{\theta_b})}$. We obtain the likelihood under the alternative hypothesis by maximising Equation~\ref{eq:incomplete} via the \textsc{em} algorithm, and the likelihood under the null hypothesis
\begin{equation}
\begin{split}
\log L_0 (\boldsymbol{\theta_a}|\boldsymbol{y}) =&~ 
\sum\limits_{i=1}^n \log f(y_i,\boldsymbol{\theta_a})
\label{eq:null}
\end{split}
\end{equation}
by analytical maximisation with respect to~$\boldsymbol{\theta_a}$.

As the single-component model is nested within the two-component model, we use their likelihood ratio to test whether the second component significantly improves the model. However, the asymptotic null distribution of the test statistic is unknown, because under the null hypothesis the mixing proportion~$\tau$ lies on the boundary of its parameter space and renders the nuisance parameter~$\boldsymbol{\theta_b}$ unidentifiable. Computationally expensive solutions are parametric bootstrapping \citep{McLachlan1987} and permutation. Briefly, we estimate the population parameter(s)~$\boldsymbol{\theta_a}$ using Equation~\ref{eq:null}, either replace the observations by simulated values from ${f(y_i,\boldsymbol{\hat{\theta}_a})}$ or permute them, fit the models from Equations~\ref{eq:incomplete} and~\ref{eq:null}, and calculate their likelihood ratio. Replacing the observations repeatedly, we obtain the empirical null distribution of the likelihood ratio test statistic. An estimate for the \mbox{$p$-value} is the proportion of test statistics that is greater than or equal to the observed one.

In preliminary simulations, parametric bootstrapping led to a higher statistical power than permutation, but also to a higher sensitivity to departures from the distributional assumption. This is possibly caused by outliers influencing the estimation of $\boldsymbol{\theta_a}$. For the mixture test, we therefore favour permutation over parametric bootstrapping. Besides, permutation is more computationally efficient and allows for permutation-based procedures to control the family-wise error rate \citep{Westfall1993}.

\subsection{Distribution of the quantitative trait}

The Gaussian mixture model (Appendix Section~C) suffers from the unbounded likelihood problem \citep{Chen2009}: if a mixture component collapses to a single observation, its mean equals this observation, and its variance equals zero. Put differently, its distribution converges to a degenerate distribution. Because the resulting likelihood function tends to infinity, the \textsc{em} algorithm has the incentive to push all but one observations in a single class. Following \citet{Chen2009}, we penalise the Gaussian likelihood function to prevent this undesirable behaviour.

The negative binomial mixture model (Appendix Section~D) suffers from its computational cost: if maximum likelihood estimation of a free parameter has no closed-form solution, numeric optimisation within each iteration of the \textsc{em} algorithm becomes necessary. Assuming a common dispersion parameter for both classes, we estimate it from the labelled observations by the maximum likelihood method. Tentatively, we allow for an offset and zero-inflation, but then estimate the mean parameters by the method of moments. Convergence to a local optimum is not guaranteed, because the moment estimates may differ from the maximum likelihood estimates.

\subsection{Genotype interactions}

In the context of \textsc{snp} effects, the mixture test is most relevant for detecting partial shifts arising from \textsc{snp}-\textsc{snp} or \textsc{snp}-environment interactions. Epistasis and genotype-environment interactions occur in a multiplicity of ways, often leading to small mixing proportions. If we are interested in the effect of the minor allele at a \textsc{snp}, the unlabelled proportion equals the minor allele frequency. Depending on the type of interaction and the allele frequencies, the effect of the minor allele might be suppressed for most and released for few individuals with the minor allele. For example, only individuals with two minor alleles at both the \textsc{snp} of interest and another \textsc{snp} might be affected. Then the mixing proportion equals the number of individuals with two minor alleles at both \textsc{snp}s divided by the number of individuals with two minor alleles at the \textsc{snp} of interest. Without knowing interacting variables, the mixture test detects \textsc{snp}s with main or interactive effects. Further tests are necessary for excluding main effects or identifying interactions.

\section{Results}

\subsection{Simulation: data generating process}

All subsequent simulation studies build upon the same data generating process:

(1)~\textit{Generating the labels.} We fix the numbers of observations $n$ and~$d$. To simulate under the null hypothesis, we set $d$ equal to zero, and to simulate under the alternative hypothesis, we set $d$ equal to a value in ${\{1,2,3,\ldots,n-2\}}$. We assign the first ${n-d}$ and the last $d$ observations to classes~A and B, respectively. This leads to the class labels $\boldsymbol{Z}={(Z_1,\ldots,Z_n)^T}$.

(2)~\textit{Simulating the observations.} We use either two Gaussian or two negative binomial distributions. The cumulative distribution functions are ${F(\boldsymbol{\cdot},\boldsymbol{\theta_a})}$ and ${F(\boldsymbol{\cdot},\boldsymbol{\theta_b})}$, where the parameter vectors $\boldsymbol{\theta_a}$ and $\boldsymbol{\theta_b}$ take different values. If sample~$i$ is in class~A (${Z_i=\mathrm{A}}$), we draw observation $Y_i$ from distribution ${F(\boldsymbol{\cdot},\boldsymbol{\theta_a})}$, and otherwise (${Z_i=\mathrm{B}}$), we draw $Y_i$ from distribution ${F(\boldsymbol{\cdot},\boldsymbol{\theta_b})}$. This leads to the observations $\boldsymbol{Y}={(Y_1,\ldots,Y_n)^T}$.

(3)~\textit{Deleting some labels.} We fix the number of unlabelled observations~$u$. The mixture test is applicable if $u$ is greater than or equal to~$d$, and $u$ is within the closed interval from $2$ to~${n-2}$. We keep the first ${n-u}$ labels, but delete the last $u$ labels. Accordingly, all labelled observations are in class~A, and $d$ out of the~$u$ unlabelled observations are in class~B.

The mixing proportion~$\tau$ is the ratio of $d$ to~$u$. After simulating the data, we use various two-sample tests to detect differences in distribution between labelled and unlabelled observations. Though, the difference lies between the two unknown classes, and not between the two known groups.

\subsection{Simulation: statistical power}

Simulating under the alternative hypothesis, we identified situations where the mixture test is more powerful than classical tests.

Under various allocations to classes (A/B) and groups (labelled/unlabelled), we simulated $100$ observations from the Gaussian distributions ${\textsc{n}(\mu_a,\sigma_a^2)}$ and ${\textsc{n}(\mu_b,\sigma_b^2)}$. Of interest are mean shifts (${\mu_a \neq \mu_b} \cap {\sigma_a^2 = \sigma_b^2}$), variance shifts (${\mu_a = \mu_b} \cap {\sigma_a^2 \neq \sigma_b^2}$), and combined shifts (${\mu_a \neq \mu_b} \cap {\sigma_a^2 \neq \sigma_b^2}$). We used the parameters ${\mu_a=0}$ and ${\sigma_a^2=1}$ for the reference distribution, and the parameters ${\mu_b \in \{0,3\}}$ and ${\sigma_b^2 \in \{1,5\}}$ for the alternative distribution. Depending on the type of shift, we compared the mixture test with the \mbox{$t$-test} of equality of means, the \mbox{$F$-test} of equality of variances, and the Kolmogorov-Smirnov test of equality of distribution.

The \mbox{$t$-test} of equality of means calculates the evidence against the null hypothesis ${H_0^*: \mu_s = \mu_u}$ in favour of the alternative hypothesis ${H_1^*: \mu_s \neq \mu_u}$, where $\mu_s$ and~$\mu_u$ are the underlying means of the labelled and unlabelled observations, respectively. The \mbox{$F$-test} test of equality of variances contrasts the hypotheses ${H_0^*: \sigma_s^2=\sigma_u^2}$ and ${H_1^*: \sigma_s^2 \neq \sigma_u^2}$, where $\sigma_s^2$ and~$\sigma_u^2$ are the underlying variances of the labelled and unlabelled observations, respectively. And the Kolmogorov-Smirnov test of equality of distribution compares ${H_0^*: F_s(\boldsymbol{\cdot}) \equiv F_u(\boldsymbol{\cdot})}$ with ${H_1^*: F_s(\boldsymbol{\cdot}) \not \equiv F_u(\boldsymbol{\cdot})}$, where ${F_s(\boldsymbol{\cdot})}$ and~${F_u(\boldsymbol{\cdot})}$ are the cumulative distribution functions of the labelled and unlabelled observations, respectively.

\begin{figure}[!htbp]
\centerline{\includegraphics[width=1.2\textwidth]{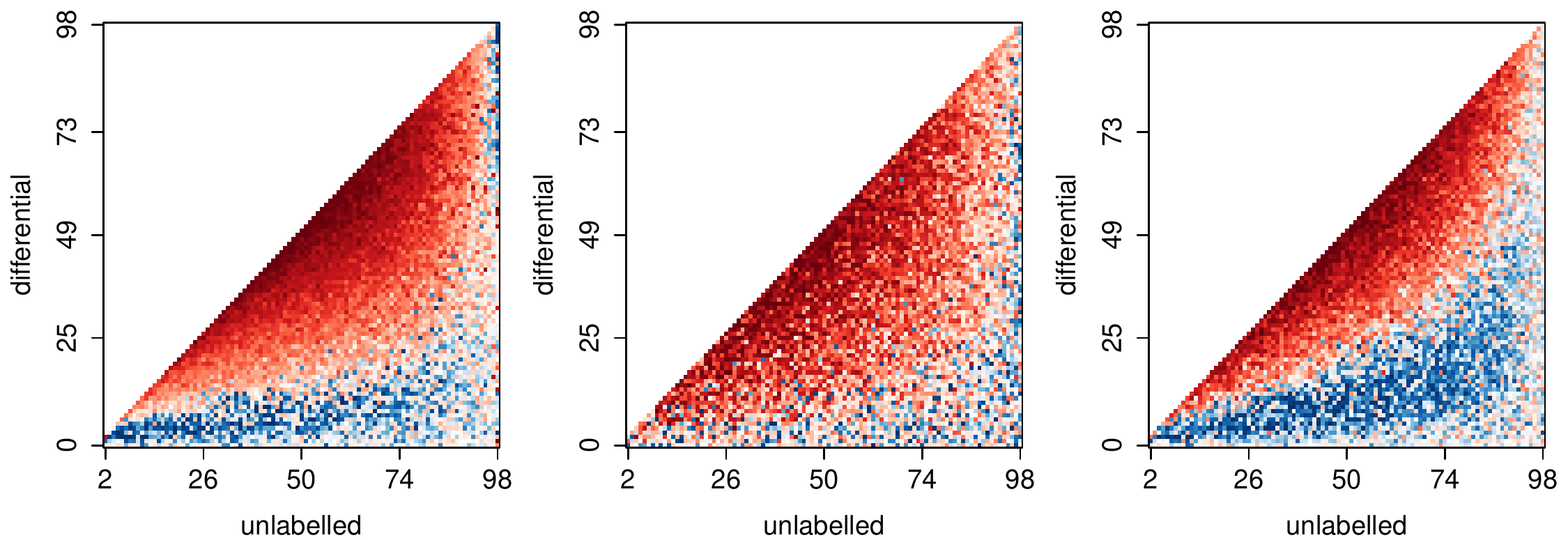}}
\caption{Ratio of \mbox{$p$-values} from the Gaussian mixture test ($100$~permutations) and the classical test, ranging from~$0.01$ (dark~blue) to infinity (dark~red), with the cutoff~one (white). The classical test is the \mbox{$t$-test} for mean shifts (left), the \mbox{$F$-test} for variance shifts (centre), or the Kolmogorov-Smirnov test for combined shifts (right). The mixing proportion is the number of observations in class~B (\mbox{$y$-axis}) divided by the number of unlabelled observations (\mbox{$x$-axis}).}
\label{fig:image}
\end{figure}

Figure~\ref{fig:image} shows the ratio of \mbox{$p$-values} from the mixture test and the classical test, with blue representing situations where the \mbox{$p$-value} from the mixture test is lower than the one from the classical test. If we increased the number of permutations, the mixture test could reach lower \mbox{$p$-values}. We conclude that ($1$)~if few unlabelled observations are in class~B, the mixture test is superior to classical tests; ($2$)~the more unlabelled observations are in class~B, the better the classical tests become relative to the mixture test; ($3$)~if all unlabelled observations are in class~B, the classical tests are superior to the mixture test.

Using the same parameters as above, we simulated ${1\,000}$ mean shifts, variance shifts, and combined shifts each, with $50$ labelled and $45$ unlabelled observation in class~A, and $5$ unlabelled observations in class~B. At the ${5\%}$~significance level, the mixture test is more powerful than the \mbox{$t$-test} for detecting mean shifts (${52\% > 17\%}$), more powerful than the Kolmogorov-Smirnov test for detecting combined shifts (${48\% > 5\%}$), but less powerful than the \mbox{$F$-test} for detecting variance shifts (${10\% < 24\%}$). This holds across all significance levels (Appendix Figure~A).

\begin{figure}[!htbp]
\centerline{\includegraphics[width=1.0\textwidth]{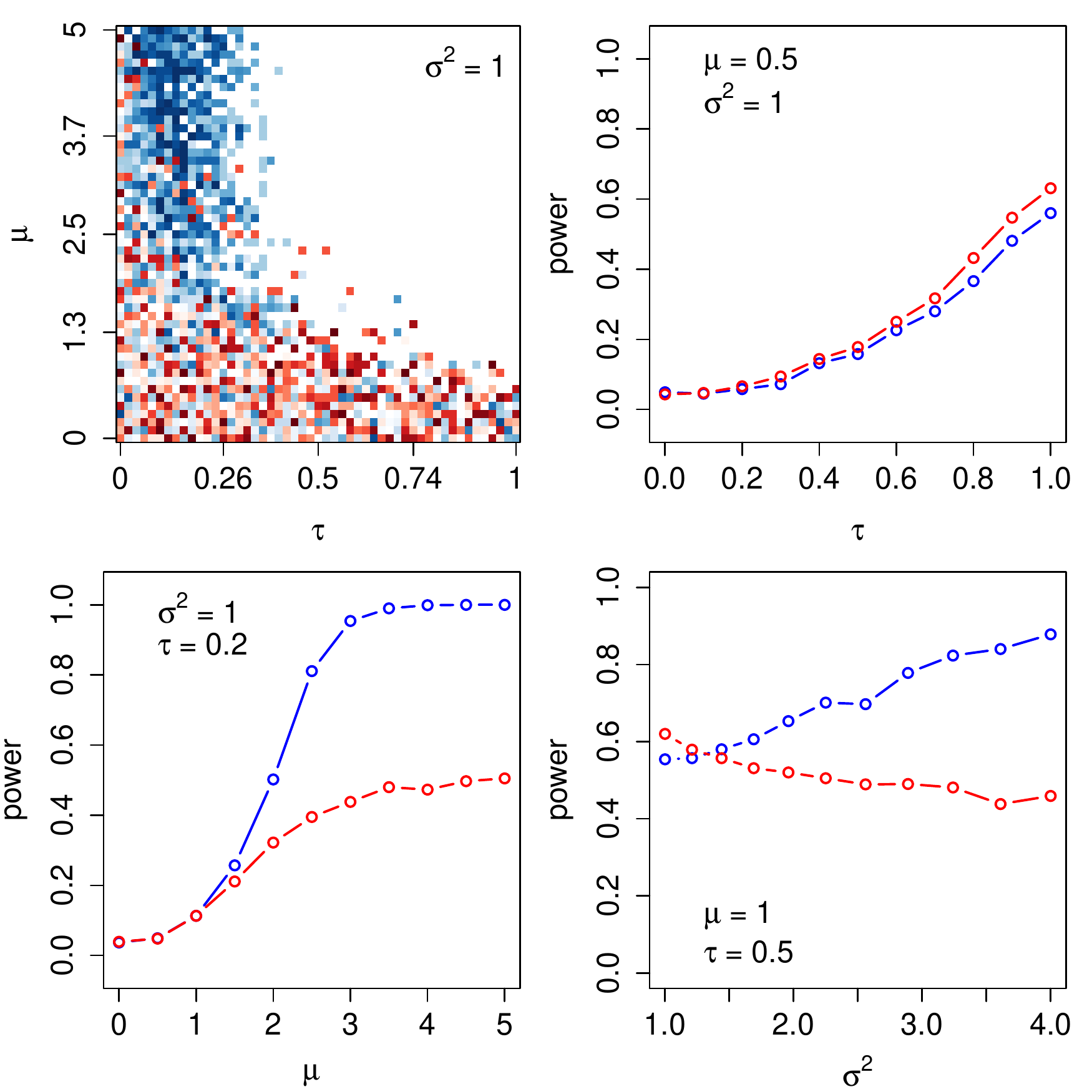}}
\caption{Simulation of $50$~labelled and ${50(1-\tau)}$~unlabelled observations from ${\textsc{n}(0,1)}$, and ${50\tau}$~unlabelled observations from ${\textsc{n}(\mu,\sigma^2)}$. The image (top left) shows the ratio of \mbox{$p$-values} from the mixture test and \texttt{PDGEtest}, ranging from~$0.01$ (dark~blue) to~$100$ (dark~red), with the cutoff~$1$ (white). The line charts show the power of the mixture test (blue) and \texttt{PDGEtest} (red) at the ${5\%}$~significance level, computed from ${1\,000}$ iterations per setting. All \mbox{$p$-values} are based on $100$~permutations.}
\label{fig:pdge}
\end{figure}

The mixture test and the non-parametric test for partial differential expression (\texttt{PDGEtest}) \citep{VanWieringen2008} test the same hypothesis on the mixing proportion. Simulating under the alternative hypothesis, we showed that the mixture test is not only more powerful at detecting combined shifts, but also at detecting mean shifts if the mixing proportion is small (Figure~\ref{fig:pdge}).

Under various allocations to classes and groups, we simulated from the negative binomial distributions ${\textsc{nb}(\mu_a,\phi)}$ and ${\textsc{nb}(\mu_b,\phi)}$, where ${\mu_a=10}$, ${\mu_b=20}$ and ${\phi=0.2}$. Assuming the dispersion parameter~$\phi$ is known, we conducted the mixture test to compare the means between classes A and~B. For comparison, we conducted three classical tests to contrast the labelled with the unlabelled populations, namely the Mann-Whitney $U$~test, the Kolmogorov-Smirnov test, and the exact test for negative binomial counts (\texttt{edgeR}) \citep{Robinson2008}. If less than half of the unlabelled observations are in class~B, the mixture test leads to lower \mbox{$p$-values} than the classical tests (Appendix Figure~B).

For mean shifts, the mixture test is much more powerful than classical tests if the mixing proportion is small, but only slightly less powerful if the mixing proportion is large (Appendix Figure~C).

One type of interactive effect does not affect any individuals in the labelled group, but does affect some individuals in the unlabelled group. Implicitly, this simulation study is about interactive effects, as labelled observations are in class~A, and unlabelled observations are in classes A and~B. In conclusion, the mixture test outperforms classical tests at detecting interactive effects.

\subsection{Simulation: false positive rate}

Simulating under the null hypothesis, we verified whether the mixture test maintains the type~I error rate.

We repeatedly generated $100$~observations from a standard Gaussian distribution, split them into a labelled and an unlabelled group, and applied the mixture test. We repeated this ${1\,000}$~times for each unlabelled percentage in ${\{5, 15, 25,\ldots,95\}}$. The false positive rate has no trend with respect to the unlabelled percentage, but fluctuates around the significance level (Appendix Figure~D).

To verify whether model misspecification renders the mixture test anti-conservative, we repeatedly generated $50$~labelled and $50$~unlabelled observations from a \mbox{$t$-distribution} with $\nu$~degrees of freedom, and applied the \textit{Gaussian} mixture test. We repeated this ${1\,000}$ times for each $\nu$ in ${\{1,2,3,\ldots,10\}}$. As $\nu$ increases, the \mbox{$t$-distribution} converges to the standard Gaussian distribution. Reassuringly, the false positive rate oscillates around the significance level across all values of~$\nu$ (Appendix Figure~D). Although the assumption of normality is not met, we do not reject too many true null hypotheses.

The negative binomial mixture test requires an estimate for the dispersion parameter. For each unlabelled percentage in ${\{5,10,15,\ldots,95\}}$, we simulated ${1\,000}$~times $100$~observations from a negative binomial distribution with ${\mu=10}$ and ${\phi=0.2}$. If we correctly estimate the dispersion parameter (${\hat{\phi}=0.2}$), the mixture test leads to a false positive rate of ${5.0\%}$ at the ${5\%}$~significance level. If we overestimate or underestimate the dispersion parameter (${\hat{\phi}=0.3}$ or ${\hat{\phi}=0.1}$), the mixture test becomes slightly anti-conservative (false positive rates ${5.2\%}$ and ${5.3\%}$).

Count data may contain excess zeros. We simulated ${1\,000}$~times $100$~observations from a zero-inflated negative binomial distribution with ${\mu=10}$, ${\phi=0.2}$ and ${\pi=0.2}$. If the underlying dispersion and zero-inflation parameters are known, the mixture test leads to a false positive rate of ${4.2\%}$ at the ${5\%}$~significance level. Overestimation of the zero-inflation parameter (${\hat{\pi}=0.4}$) keeps the mixture test conservative (false positive rate=${4.6\%}$), but underestimation (${\hat{\pi}=0}$) renders the mixture test anti-conservative (false positive rate=${6.9\%}$). In practice, $\phi$ and $\pi$ are unknown. Under a sample size of ${1\,000}$, maximum likelihood estimation leads to a slightly anti-conservative test (false positive rate=${6.0\%}$).

Conclusively, the mixture test maintains the type~I error rate if it is correctly specified. To achieve this, the chosen distribution should match the distribution of the quantitative trait. Quantile-quantile plots and cumulative distribution plots of the labelled observations can help to make the right choice.

Furthermore, the validity of the negative binomial mixture test depends on the reliability of the dispersion estimate. If the number of labelled observations is large, maximum likelihood estimation is reliable. If the sample size or the labelled proportion is small, shrinkage estimation should be used. In the presence of excess zeros, the zero-inflated negative binomial mixture test might be more appropriate.

\subsection{Application: GWAS}

In this application, we used the semi-supervised mixture model to conduct a genome-wide association study~(\textsc{gwas}).

Analysing data from the Longitudinal Aging Study Amsterdam \citep{Huisman2011}, our aim is to detect \textsc{snp}s significantly associated with the body mass index~(\textsc{bmi}). \textsc{snp} data measured with an exome array are available for ${n=847}$~individuals and ${242\,857}$~loci. We transformed them to binary covariates, such that one level of a covariate represents zero minor alleles, and the other level represents one or two minor alleles at the corresponding locus. Excluding all loci with fewer than $50$~individuals on either level, or with missing values, leads to ${p=26\,517}$~\textsc{snp}s on the autosomes.

To detect \textsc{snp}s associated with sex, we constructed one contingency table for each \textsc{snp} (male/female and labelled/unlabelled), and applied Pearson's chi-squared test of independence. At the Bonferroni-adjusted ${5\%}$~level, the only significant \textsc{snp} is \textit{exm2277017}. Whereas all $425$~females have zero minor alleles at this \textsc{snp}, $414$~out of $422$ males have one or two minor alleles. Questioning the quality of the genotyping, we excluded this sex-biased genotyping probe from the analysis. 

According to the \mbox{$t$-test} of equality of means and the \mbox{$F$-test} of equality of variances, the \textsc{bmi} differs significantly in mean and variance between males and females (Appendix Figure~E). To adjust for obvious confounding variables, we linearly regressed the logarithmic \textsc{bmi} on sex and age. We observe the residuals approximately follow a Gaussian distribution (Appendix Figure~F). This qualifies them as a response variable for the semi-supervised mixture model. For one \textsc{snp} at a time, individuals with zero minor alleles belong to the first mixture component, and those with one or two minor alleles belong to either component. Using the semi-supervised mixture model, we test whether the two components are significantly different from each other.

\begin{figure}[!htbp]
\centerline{\includegraphics[width=1.0\textwidth]{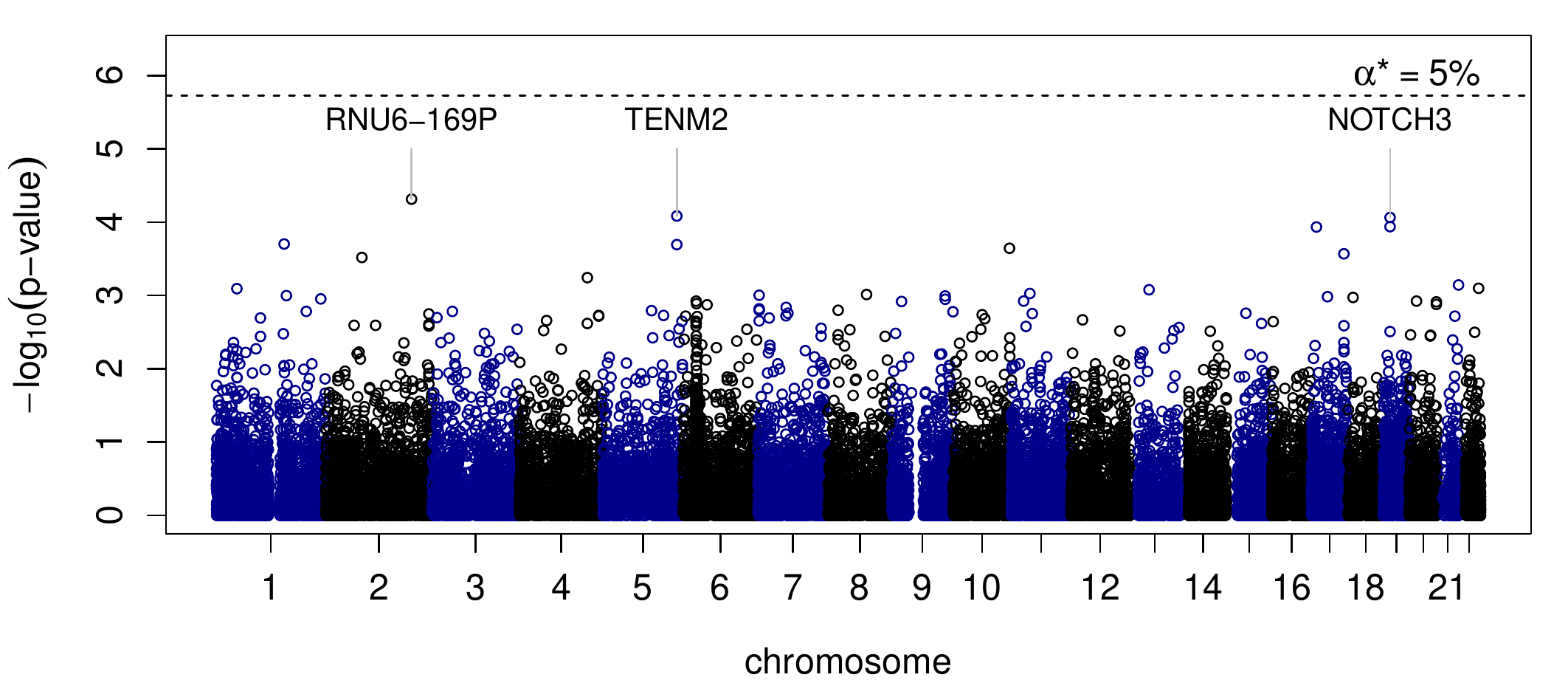}}
\caption{Manhattan plot showing \mbox{$p$-values} and genomic locations of \textsc{snp}s. The dashed line represents the threshold for Bonferroni significance at the ${5\%}$~level. The labels indicate the nearest genes to the top three most significant \textsc{snp}s.}
\label{fig:manhattan}
\end{figure}

Applying permutation testing to all \textsc{snp}s is computationally expensive, since we need more than half a million iterations to make Bonferroni-corrected \mbox{$p$-values} below the ${5\%}$~significance level possible. Setting the maximum number of iterations to one million, we repeatedly interrupted iteration, and only continued with \textsc{snp}s that still had a chance of reaching a low \mbox{$p$-value}. Specifically, a \textsc{snp} was dropped as soon as more than $10$ of its simulated test statistics were larger than or equal to its observed one. Already after ${1\,000}$ iterations, the active set comprised no more than $238$~\textsc{snp}s, and only $18$~\textsc{snp}s reached ${10\,000}$~iterations. The top three most significant \textsc{snp}s are \textit{exm2269357} close to gene \textit{RNU6-169P} on chromosome~$2$, \textit{exm501828} in gene \textit{TENM2} on chromosome~$5$, and \textit{exm1436489} in gene \textit{NOTCH3} on chromosome~$19$. All three \textsc{snp}s lead to partial shifts between the two groups (Appendix Figure~G). \cite{Bryan2014} have previously reported the \textsc{bmi}-susceptibility of \textit{TENM2}. Figure~\ref{fig:manhattan} shows all \mbox{$p$-values} and their genomic locations.

\begin{figure}[!htbp]
\centerline{\includegraphics[width=1.0\textwidth]{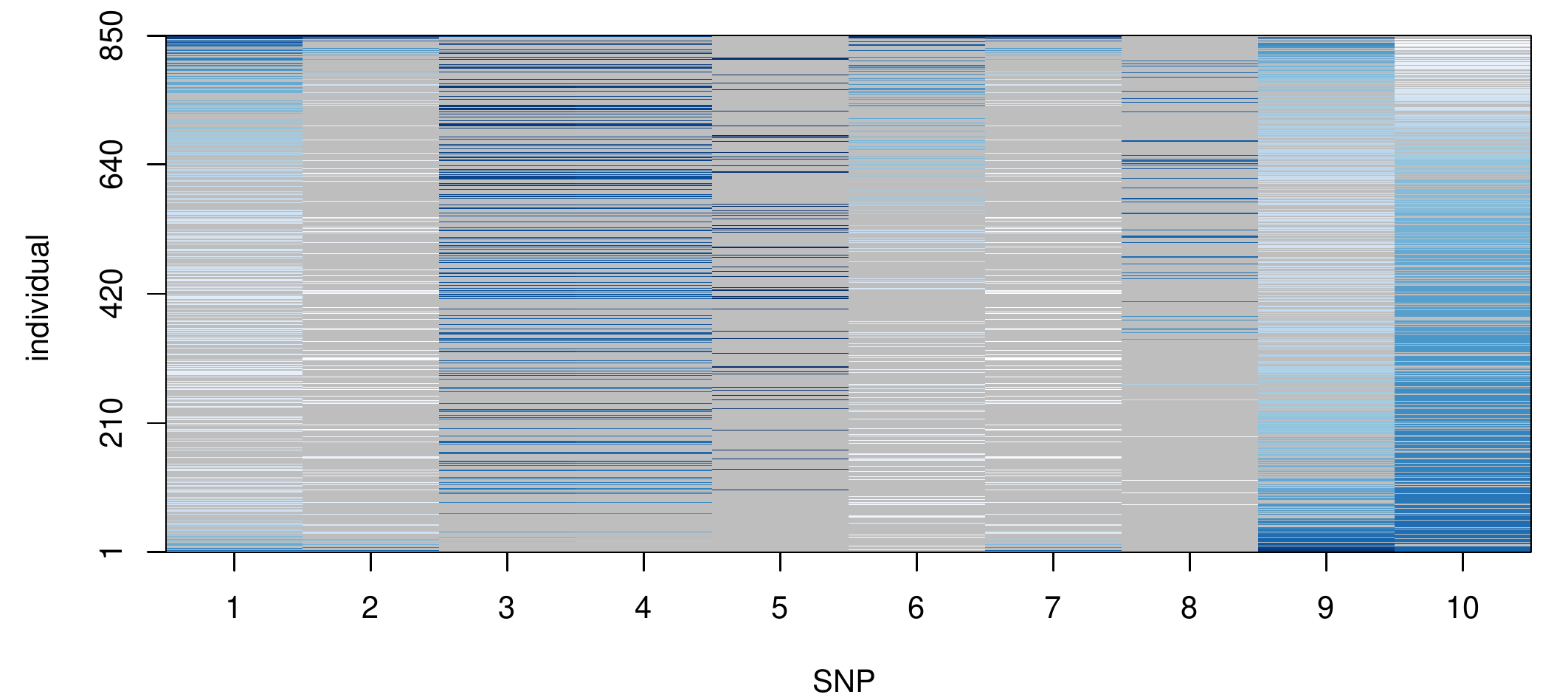}}
\caption{Membership probabilities for labelled observations equal zero (grey), and those for unlabelled observations range from zero (white) to one (dark~blue). The top ten most significant \textsc{snp}s (\mbox{$x$-axis}) are ascendingly ordered by their \mbox{$p$-value}, and the individuals (\mbox{$y$-axis}) are ascendingly ordered by their quantitative trait.}
\label{fig:posterior}
\end{figure}

Figure~\ref{fig:posterior} shows the membership probabilities for the most significant \textsc{snp}s and all individuals. By construction, labelled observations have a membership probability of zero. At the most and second-most significant \textsc{snp}s (\textit{exm2269357}, \textit{exm501828}), merely ${47\%}$ and ${25\%}$ of the unlabelled observations have a membership probability above~$0.5$, suggesting main or interactive effects. At the third-most significant \textsc{snp} (\textit{exm1436489}), all unlabelled observations have a membership probability close to one, suggesting a main effect. However, all three effects are insignificant at the Bonferroni-adjusted ${5\%}$~level.

Several tests could examine whether the adjusted \textsc{bmi} differs significantly between individuals with zero minor alleles at a \textsc{snp}, and those with one or two minor alleles. Exploiting the Gaussian distribution, we used the \mbox{$t$-test} of equality of means. The most significant \textsc{snp} is \textit{exm1436475} in gene \textit{NOTCH3} on chromosome~$19$, which is again insignificant at a Bonferroni-controlled family-wise error rate of~${5\%}$.

Hence, we fail to identify significant associations between the adjusted \textsc{bmi} and \textsc{snp}s. We suspect the sample size is too small and the signal is too weak for reaching significance. For example, in \citet{Robinson2015}, \citet{Speliotes2010} and \citet{Yang2015}, where significant \textsc{snp}s are found, the sample sizes are between one and three orders of magnitude larger.

\subsection{Application: eQTLs}

In this application, we used the semi-supervised mixture model to detect expression quantitative trait loci~(e\textsc{qtl}s).

Analysing data from \citet{Lappalainen2013}, our aim is to detect \textsc{snp}s significantly associated with \mbox{\textsc{rna-s}eq} gene expression. Both \textsc{snp} data and \mbox{\textsc{rna-s}eq} data are available for $373$~individuals of European ancestry. We excluded all loci with a minor allele frequency below~$5\%$, calculated the absolute Pearson correlation coefficients for all pairwise combinations of \textsc{snp}s within the same gene, and repeatedly removed the most correlated \textsc{snp} until decreasing all coefficients to below $0.5$. We excluded all noncoding genes, and adjusted for different library sizes using the trimmed mean normalisation method (\texttt{edgeR}) \citep{Robinson2010}. This leads to ${144\,774}$ pairs consisting of a gene and a \textsc{snp} within the gene.

Analysing one pair at a time, we assume \mbox{\textsc{rna-s}eq} gene expression follows a negative binomial distribution, or a mixture of two negative binomial distributions with equal dispersion parameters. Individuals with zero minor alleles are in the labelled group, and those with one or two minor alleles are in the unlabelled group. After estimating the dispersion parameters by the maximum likelihood method, we conducted the mixture test. Given a gene with $p$~\textsc{snp}s, we need at least ${p/0.05}$~iterations to make significant \mbox{$p$-values} at the Bonferroni-adjusted ${5\%}$~level possible. We set the maximum number of iterations to this limit, and interrupted iteration as soon as it became impossible to reach significance. Genome-wide instead of gene-wide significance would require many more permutations. For comparison, we applied the Mann-Whitney $U$~test, and the exact test for negative binomial counts (\texttt{edgeR}) \citep{Robinson2008}, examining whether \mbox{\textsc{rna-s}eq} gene expression differs between labelled and unlabelled individuals.

We multiplied the raw \mbox{$p$-values} by the number of \textsc{snp}s within the corresponding gene, and used the ${5\%}$ significance level. Here, the exact test detects slightly more significant associations than the Mann-Whitney $U$~test. Among all pairs, the mixture test and the Mann-Whitney $U$~test detect ${2\,651}$ and ${3\,341}$ significant associations, respectively, with an overlap of~${2\,055}$. We are interested in the $596$ pairs receiving significant \mbox{$p$-values} from the mixture test, but insignificant \mbox{$p$-values} from the Mann-Whitney $U$~test. The corresponding \textsc{snp}s potentially affect gene expression through interactions.

For each pair, we compared the distribution of \mbox{\textsc{rna-s}eq} gene expression between the labelled and the unlabelled group. Pairs with significant \mbox{$p$-values} from both tests and those with only a significant Wilcoxon \mbox{$p$-value} have larger differences than those with only a significant mixture \mbox{$p$-value} (Appendix Figure~H). This indicates the mixture test detects partial differential expression. As by-products, the mixture test estimates the mixing proportion and the difference between the class means (Appendix Figure~I). If both are large, the mixture test and the Mann-Whitney $U$~test are likely to detect a significant association. The mixture test is superior to the Mann-Whitney $U$~test if the mixing proportion is small, and inferior if the mixing proportion is large.

For some randomly selected pairs, we checked the distributional assumption. Examining one pair at a time, we adjusted for different library sizes, fitted a negative binomial distribution to the labelled group, used the mean and dispersion estimates to simulate ${1\,000}$~expression values for each labelled individual, and compared the observed to the simulated data. The quantile-quantile plots show strong departures from the negative binomial assumption for some pairs, also due to zero-inflation (Appendix Figure~K). For the same pairs, the approximate Kolmogorov-Smirnov test rejects the distributional assumption at the ${5\%}$~significance level.

Consequently, we also conducted the zero-inflated negative binomial mixture test. Using the maximum likelihood method, we simultaneously estimated the dispersion and zero-inflation parameters. In around ${52\%}$ of the pairs, the zero-inflation parameter equals zero, rendering the models with and without zero-inflation equivalent. In around ${48\%}$ of the pairs, both the zero-inflation and the dispersion parameters account for the variability in the data. Subsequently, if the zero-inflation estimate is greater than zero, the dispersion estimate from the zero-inflated model is lower than the one from the non-inflated model (Appendix Figure~J). The number of significant \mbox{$p$-values} decreases from ${2\,651}$ to ${2\,578}$.

Hence, the mixture test and the Mann-Whitney $U$~test detect partially overlapping sets of significant associations. Because each test detects numerous e\textsc{qtl}s the other test misses, they complement each other. Integrated into a two-stage analysis, the mixture test could help to identify significant \textsc{snp}-\textsc{snp} interactions.

\section{Discussion}
\label{discussion}

We have proposed a semi-supervised mixture test to detect effects of \textsc{snp}s on a quantitative trait. We have shown by simulation and application that it detects meaningful signals.

Based on the number of minor alleles, we allocate individuals to two groups. We either combine zero with one minor alleles, or one with two minor alleles. If the minor allele frequency is low, only the latter choice leads to sensible group sizes. Thereafter, we test for a mixture distribution in the group including zero minor alleles, or in the group including two minor alleles. In short, the so-called modification may represent the presence or absence of the minor or major allele. Instead of proceeding similarly for all \textsc{snp}s, we could establish a decision rule depending on the allele frequencies. Potentially, the test can be extended to mixture models with more than two groups, or with more than two classes, accounting for the effects of heterozygous genotypes and multiple \textsc{snp}s.

We have implemented the semi-supervised mixture test for the Gaussian and the negative binomial distributions, allowing for a wide variety of quantitative traits. Although extensions to other distributions are conceptually simple, they are computationally prohibitive in the case of numeric optimisation within the \textsc{em} algorithm. Even for distributions with closed-form estimates, permutation remains too computationally expensive for reaching Bonferroni-significance in high-dimensional settings. We increased computational efficiency by interrupting resampling when it becomes impossible to reach the significance level within the maximum number of permutations. If the null distribution of the test statistic was known, \mbox{$p$-values} could have been obtained instantaneously. As a workaround, we might first obtain \mbox{$p$-values} from an approximate null distribution, and then apply permutation to the most promising \textsc{snp}s.

Crucially, the mixture test detects \textsc{snp}s with main or interactive effects, without detecting the other interacting variables. Although the membership probabilities may suggest potential interacting variables, this is exploratory in nature. To test \textsc{snp}-\textsc{snp} or \textsc{snp}-environment interactions, the mixture test can be integrated into a multi-stage testing procedure. One option is to split the individuals into two independent sets, conduct the mixture test in the first set, identify potential interactions, and test them in the second set \citep{Pecanka2017}. Another option is to devise a hierarchy of tests, and to correct for multiple testing at each level of the hierarchy \citep{Meinshausen2008}. In both cases, the multiple testing correction for pairwise interactions decreases considerably.

The semi-supervised mixture model and test can do more than analysing \textsc{snp}s: they are applicable whenever effects of a binary variable on a numerical variable are of interest.

\section*{Software}

The R package \href{https://doi.org/10.18129/B9.bioc.semisup}{\texttt{semisup}} runs on any operating system equipped with R-3.4.0 or later. It is available from Bioconductor under a free software license: \newline http://bioconductor.org/packages/semisup/.


\vspace{1cm}

\textbf{Supplementary material:} The appendix including mathematical details and additional figures is available upon request.

\vspace{0.5cm}

\textbf{Authors' contributions:} Based on an idea from MAJ, AR implemented the method and drafted the manuscript. NMvS provided the data. RXM, MAvdW and MAJ revised the manuscript critically. All authors read and approved of the final manuscript.

\vspace{0.5cm}

\textbf{Acknowledgements:} This research was funded by the Department of Epidemiology and Biostatistics, VU University Medical Center Amsterdam. The Longitudinal Aging Study Amsterdam is supported by a grant from the Netherlands Ministry of Health Welfare and Sports, Directorate of Long-Term Care.


\bibliographystyle{apalike}
\bibliography{bibliography}

\end{document}